\documentclass[a4paper]{article}

\usepackage{graphicx}
\usepackage{dcolumn}
\usepackage{bm}
\usepackage{hyperref}
\hypersetup{hidelinks}
\usepackage{subcaption} 
\usepackage[round]{natbib}
\usepackage{DirkMacros}
\usepackage{siunitx}
\usepackage[nameinlink,noabbrev,capitalise]{cleveref}
\usepackage{xcolor}
\usepackage{nicefrac}
\usepackage{ragged2e}
\usepackage{caption}
\usepackage{amssymb}
\usepackage[margin=1in]{geometry}
\DeclareCaptionJustification{justified}{\justifying}
\title{Reversal of a flat plate into its wake: \\a minimal model for wake capture}

\author{Dirk de Boer\textsuperscript{1}, Abel-John Buchner\textsuperscript{1}\\ \\
\small \textsuperscript{1}Delft University of Technology, Faculty of Mechanical Engineering, \\ \small Delft, Zuid-Holland, The Netherlands\\
\small Corresponding author: a.j.buchner@tudelft.nl}

\date{}

\begin{document}
\maketitle

\begin{abstract}
In reciprocating motions, such as those observed in insect flight, wing--wake interaction is known to play a crucial role in fluid force generation. While the existence of this effect has been acknowledged, particularly in explaining discrepancies between measured forces and quasi-steady approximations, fundamental research on the mechanism underlying this interaction and its scaling remains limited. To address this gap, this study investigates the excess drag force, relative to quasi-steady estimates, acting on a flat plate during the reversal phase of a forward and back translational motion along a path normal to the plate's surface. The flow produced by this motion, studied at Reynolds numbers relevant to insect flight, serves as a simplified analogue to biological flapping. It is demonstrated that interaction during reversal with pre-existing wake structures indeed generates drag in excess of quasi-steady predictions. The primary parameter governing this interaction is the distance travelled before reversal, which not only influences the magnitude of the peak additional drag but also its temporal dynamics. We hypothesise that these observations are closely linked to optimal vortex formation, as the time trace of the additional drag during reversal is qualitatively altered by the detachment of the starting vortex ring and subsequent formation of a new vortex ring. Specifically, while the peak wake force increases with increasing vortex ring strength, vortex detachment and subsequent formation lead to the appearance of two distinct wake-force peaks. Furthermore, as the pre-reversal distance traversed increases, the wake interaction force post-reversal decays more slowly. Observations of the corresponding flow fields reveal a similar spatial decay of the streamwise velocity in the wake at the moment of reversal, suggesting a direct link.   Representing the starting vortex as a point vortex captures the spatial scaling of the wake over a range of plate traversal distances, indicating that the spatial scaling of the wake, and commensurately the temporal scaling of the wing--wake interaction effect, is primarily governed by the vortex ring’s position, shape, and circulation. This simplification reveals that the dependence on the pre-reversal translation distance can be described by a combined fourth-root and linear scaling.
\end{abstract}

\section{Introduction}

Unsteadiness is inherent to many natural and engineered flows, resulting in complex flow phenomena and force-generation mechanisms. Examples include impulsive flows found in rowing~\citep{Grift_2019}, automotive racing~\citep{Zhou_2026}, and offshore structures in waves~\citep{Lightill_1986}. A setting in nature where unsteady mechanisms play a role, and which has been extensively studied over the past decades, is insect flight. Through unsteady aerodynamic mechanisms, insects are able to delay or avoid stall at high angles of attack, and generate lift beyond steady-state limits~\citep{Ellington_1984b, Ellington_1996, Mulleners_2013}. Understanding these mechanisms is therefore essential not only for explaining biological flight, but also for uncovering principles that can be translated to the design of bio-inspired technologies. 

Numerous efforts have been made to decompose, interpret, and understand the source of flow-generated forces in unsteady flows. One method of doing so is to compute forces directly from flow field data, invoking the governing equations of fluid flow in methods involving momentum conservation~\citep{Noca_1997, Rival_2017, Limacher_2020} or vorticity-related constructions~\citep{Menon_2021, Gehlert_2023, Prakhar_2025, Otomo_2025, Savelli_2026}. However, these approaches rely on the availability of flow field information, and therefore lack predictive capability from kinematics alone, serving instead as primarily a measurement or analytical tool.    

In the context of insect flight, quasi-steady (QS) models are commonly used to predict fluid forces purely from flapping kinematics~\citep{Ellington_1984a, Dudley_1990, Sane_2002, Dickinson_2016, Cai_2021, VanVeen_2022, Wang_2016}. These models decompose flow-generated forces into several components associated with distinct physical phenomena. One of these components is the translational force, which scales with the square of the body's velocity. Another component is the added-mass force, arising from potential flow theory and representing the inertia of the fluid accelerated with the body. The acceleration of the body modifies the flow potential and generates a pressure difference, resulting in a force proportional to the body's acceleration. Additional components account for the forces on the body caused by its rotation. In this framework, the individual components are assumed to depend solely on the instantaneous kinematics, with the total force obtained through their linear addition.

Although QS models often provide reasonable force estimates~\citep{Sane_2002}, there are limitations to the approach. Typically, the added mass is treated as constant, with estimates available for various geometries~\citep{Yu_1945, Patton_1965, Payne_1981}. However, recent studies on accelerating plates have demonstrated that the classical constant-added-mass formulation, despite being grounded in potential flow theory, fails to accurately capture the forces associated with accelerations of non-negligible duration~\citep{Grift_2019}. This motivated the development of an alternative scaling law~\citep{Reijtenbagh_2023, Reijtenbagh_2026}, in which the added mass contribution is reformulated and persists beyond the acceleration phase as a decaying history force. The physical origin of this force has been attributed to the generation and subsequent diffusion of vorticity produced at the body surface during acceleration, introducing a dependence on kinematic history. Additionally, bodies moving at a constant velocity after acceleration experience a start-up transient due to wake development, known as the Wagner effect~\citep{Wagner_1925}. This effect, described by the empirical Wagner function, characterizes the gradual build-up of fluid forces from zero to their steady-state values as the starting vortices, shed during initial acceleration, advect downstream toward infinity. This indicates that, due to kinematic time history effects, forces cannot be presumed constant, even during constant-velocity translation. Some QS formulations therefore include this effect as a correction~\citep{Wang_2016, VanVeen_2022}. 

Another weakness of QS models is their inherent ignorance of existing flow structure, which may interact with the body. In systems involving kinematic reciprocation, the body passes through its own wake region and interacts with the flow structures in the wake. Specifically, in periodic flapping motions in biology, wings repeatedly interact with the wake vortices generated by previous strokes. This wing--wake interaction has the potential to significantly influence the generation of fluid-dynamic forces~\citep{Dickinson_1994}.

As the wing interacts with previously shed wake structures, the resulting force generation depends on the state of the wake at the moment the wing encounters it. Understanding these interactions therefore requires an understanding of how wake structures are formed and evolve. The no-slip condition at solid boundaries generates strong velocity gradients near body edges, producing vorticity~\citep{Saffman_1992}. This vorticity is advected from the body's edges to form separated shear layers, which subsequently roll up into spatially-compact vortical structures, typically in the form of three-dimensional vortex rings~\citep{Von_Ellenrieder_2003, deGuyon_2021}. These vortex rings can only accumulate circulation up to a limiting value~\citep{Gharib_1998, Dabiri_2005, Sun_2025}, after which they detach from the body in a process known as pinch-off. Pinch-off is, if the body's motion continues, often followed by the re-formation of subsequent vortices. Consequently, the wing--wake interaction force is inherently linked to the wing kinematics and its time history. Models accounting for such history-dependent hydrodynamic forces have also been developed for other flow regimes under specific assumptions. A classical example is the Basset history force appearing in the Basset--Boussinesq--Oseen equation \citep{Brennen_2005} for unsteady flow at very low Reynolds number, as well as its later generalisation in the Maxey--Riley equation~\citep{Maxey_1983}. The Basset history force originates from the delayed response of the viscous flow during unsteady motion, taking the form of a convolution integral over the acceleration history with a memory kernel that arises from the diffusive growth of the vorticity boundary layer~\citep{Basset_1888}. 

Analogously to the Basset history force, one can attempt to consider the force generated by wing--wake interaction also at higher Reynolds numbers as an additive time-history-dependent term. In the literature on wing--wake interaction, two main approaches can be distinguished in isolating such a term. The first compares the forces computed or measured during a given flapping cycle with those from the initial half-cycle in a series, and attributes the difference to wing--wake interaction~\citep{Lua_2017, Lee_2018, Li_2022, Li_2024}. The underlying assumption is that the initial half-cycle is free of wing--wake interaction, since no wake has yet been generated. The second approach employs a QS model, and attributes the difference between the total computed or measured aerodynamic force and the QS prediction to wing--wake interaction. This approach assumes that the QS model accurately captures all force contributions that depend solely on the instantaneous kinematics, whereas wing--wake interaction represents a history-dependent effect~\citep{Dickinson_1999, Sane_2002, Nakata_2015, Bomphrey_2017}. 

Application of the aforementioned approaches has led to a broad range of findings regarding wing--wake interaction. Most studies have focused on its influence on lift generation, although a limited number have also examined its effect on drag~\citep{Lua_2017, Lee_2018}. Several studies report that wing--wake interaction enhances lift generation~\citep{Lua_2017, Bomphrey_2017, Lee_2018}. In contrast, others show that its effect is strongly dependent on the wing kinematics, with lift either increasing or decreasing depending on the timing of wing rotation~\citep{Dickinson_1999}, the pitching kinematics~\citep{Li_2022}, or three-dimensional vortex dynamics~\citep{Li_2024}. The wing--wake interaction phenomenon is, in this context, commonly referred to as ``wake capture'', reflecting that part of the energy stored in the wake is recovered and converted into aerodynamic force~\citep{Sane_2003, Lehman_2008}. Wing--wake interaction is believed to play a particularly important role in mosquito hovering flight because of the mosquitoes' characteristic combination of low stroke amplitudes and high flapping frequencies~\citep{Bomphrey_2017, Liu_2020, LeRoy_2026}.

Although a wide range of studies on wing--wake interaction exists, no general consensus has yet emerged regarding the underlying mechanisms and scaling laws governing its aerodynamic effects. Consequently, a more fundamental understanding of wing--wake interaction is required. To this end, the present work considers the flow around a flat plate undergoing a single degree-of-freedom (DOF) translation normal to its surface followed by a reversal. This kinematic configuration is chosen because it represents the simplest body--wake interaction, serving as a reduced analogue of biological flapping. Related canonical interaction problems have been studied extensively, including gust encounters with lifting surfaces or plates~\citep{Amiet_1990, Ramamurti_2008, Viswanath_2010, Perotta_2017, Hufstedler_2019, Biler_2019} and the impingement of vortex structures on solid or porous boundaries~\citep{Lim_1991, Chu_1993, Adhikari_2009, McAtee_2026}. However, these configurations differ fundamentally from the interaction between a moving body and its own wake, and their primary focus is generally on the qualitative evolution of the interacting flow structures rather than on quantifying and scaling the forces induced by the interaction. For the present kinematics, the symmetry of both the flow field and the plate geometry causes the lift force to vanish, leaving only drag generation to be considered. The objective of this work is to determine how interaction with previously shed wake structures influences the drag force experienced by a flat plate during reversal.

To address this objective, numerical simulations are performed and validated experimentally. The performed kinematics, numerical methodology, and QS model used to isolate wake interaction effects are described in~\cref{sec:Method}. Results are presented in~\cref{sec:Results}, followed by the conclusions in~\cref{sec:Conclusion}. Convergence and experimental validation of the numerical simulations are provided in~\cref{sec:appendix_convergence} and~\cref{sec:appendix_expval}, respectively.

\section{Method}
\label{sec:Method}

\subsection{Kinematics \& Relevant Parameters}

\begin{figure}
    \subcaptionbox{\centering\hspace{-1.25cm}\label{fig:schematic_kinematics}}{\includegraphics[width=0.5\textwidth]{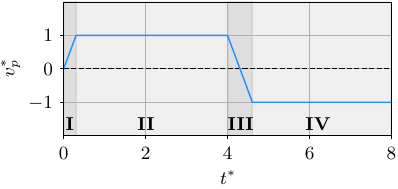}}%
    \subcaptionbox{\centering\hspace{1.4cm}\label{fig:schematic_motion}}{\includegraphics[width=0.5\textwidth]{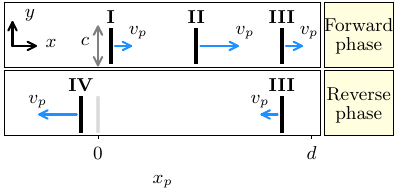}}%
    \caption{Kinematics of the flat plate forward translation and reversal motion studied here. (a) Dimensionless velocity, $\vpstar$, over time, for a case reversing at $\dstar=4$. (b) Schematic of indicative plate location and motion direction during selected phases of motion. Forward and reverse phases of motion are vertically offset for ease of visualisation. Roman numerals delineate distinct phases of the motion history: \textbf{I}. linear acceleration, \textbf{II}. translation at constant forward velocity, $\vpstar$, in $\xstar$, \textbf{III}. linear deceleration, \textbf{IV}. translation at constant reverse velocity, $-\vpstar$.}
\label{fig:schematic_motion_cycle}
\end{figure}

The force generated by wing--wake interaction is studied for a thin flat plate with chord length, $c$, and span, $b$, undergoing a single-degree-of-freedom (DOF) translation normal to its surface, followed by a reversal in the opposite direction. The plate position is denoted by $\xp$. Starting from rest, the plate accelerates briefly with a constant acceleration $\accp$, translates at a constant velocity $\vp$, and subsequently decelerates with acceleration $-\accp$ until reaching an equal but opposite velocity $-\vp$ (\cref{fig:schematic_motion_cycle}). Such translation-then-reversal motion constitutes the kinematically simplest form of body--wake interaction and serves here as a baseline for isolating this effect. The acceleration $\accp$ is set to a finite value in this study, although the effect of acceleration is itself not the focus here. To clearly distinguish the wake interaction force from acceleratory effects, the time spent accelerating is minimised by setting $\accp$ as high as experimentally feasible. The plate aspect ratio, defined as $AR=b/c$, is fixed at $AR=4$. This value is motivated by the wing aspect ratio of the mosquito species \textit{Aedes aegypti}, and retains an appreciable degree of three-dimensionality in the resulting flow structures. The plate thickness is $c/15$. The flat-plate geometry reduces problem complexity by fixing the flow separation point while effectively approximating a flapping-wing cross-section. 

The primary parameter considered in this study is the nondimensional distance, $\dstar=d/c$, travelled prior to reversal, where $d$ denotes the plate position $\xp$ at the moment of reversal. This parameter determines the state of wake development encountered by the plate during reversal and is therefore expected to strongly influence both the magnitude and temporal evolution of the wake interaction force. To investigate the influence of wake development, the pre-reversal distance is varied over the range $\dstar\in [1,8]$. This range encompasses interactions with the initial starting vortex ring at low $\dstar$ and with more developed wakes containing pinched-off and re-formed vortices at larger $\dstar$, including the expected optimal vortex formation length near $\dstar\approx4$~\citep{Gharib_1998}. Two Reynolds numbers, defined as $\Rey=\vp c/\nu$, with $\nu$ being the kinematic viscosity, are considered within the insect-relevant range, namely $\Rey=100$ and $\Rey=1000$, allowing the influence of viscous effects on wake development and wing--wake interaction to be investigated.

\subsection{Numerical Simulation}

To clearly distinguish wake interaction forces, clean force data are obtained using the Julia-based~\citep{Bezanson_2017} solver, \textit{WaterLily.jl}~\citep{Weymouth_WL_2025}. It numerically solves the incompressible Navier-Stokes equations, enforcing boundary conditions on a static grid using the boundary data immersion method~\citep{Maertens_2015}. Although the physical problem involves a translating plate, the plate remains stationary in the computational domain. Instead, the prescribed motion is imposed by time-dependent inflow boundary conditions, yielding an equivalent moving-flow formulation. The timestep is dynamically adjusted to maintain a prescribed maximum Courant--Friedrichs--Lewy (CFL) number, ensuring numerical stability throughout the simulations. \textit{WaterLily.jl} has been chosen for its ease of implementation and speed, particularly because it does not require re-meshing for moving bodies. Previous applications of \textit{WaterLily.jl} include free-falling objects~\citep{Weymouth_2026} and fish propulsion~\citep{Zhu_2026}. 

The solver uses a uniform grid spacing throughout the numerical domain. While this facilitates the numerical implementation, it requires the far field to be resolved at the same spatial resolution as the vicinity of the body, resulting in substantial memory requirements for large computational domains. To mitigate this, a Biot--Savart-based boundary condition (BC)~\citep{Weymouth_Biot_2025} is implemented, which computes the velocity at each boundary point from the vorticity field in the interior of the numerical domain according to the Biot--Savart law. Provided that no vortical structures cross the numerical boundaries, this approach effectively reproduces an unbounded flow field while substantially reducing the required extent of the computational domain. 

To determine suitable spatial discretization and domain dimensions, a convergence study was performed. The study was conducted for the most demanding case, corresponding to the highest Reynolds number and the shortest pre-reversal distance, as increasing $\Rey$ and decreasing $\dstar$ both lead to smaller flow structures requiring finer spatial resolution. Grid convergence tests showed that a resolution of $64$ grid cells per chord length $c$ is sufficient, resulting in typical force differences of less than $3\%$ compared with a discretization of $128$ cells per chord. Subsequently, the use of two symmetry planes ($xy$ and $xz$) to reduce the numerical domain by a factor of four was validated by comparison with a simulation of the full computational domain, yielding force differences below $2\%$. Finally, the computational domain size was varied, demonstrating that the computed forces converge to within $1\%$ provided that vortical structures remain within the domain boundaries. For the case with $\Rey=1000$ and $\dstar=1$, this resulted in a computational domain of dimensions $(\xstar,\ystar,\zstar)=(5,2,3.5)$, with the plate positioned such that one quarter lies within the simulated domain due to the imposed symmetry conditions. Higher $\dstar$ cases required an extension of the computational domain in both the direction of plate motion ($x$) and along the plate chord ($y$), to accommodate the increased plate displacement and the growth of the vortex ring, respectively. Details of the convergence study and experimental validation are provided in~\cref{sec:appendix_convergence,sec:appendix_expval}, respectively.

\subsection{Quasi-Steady prediction}

By definition, QS models depend solely on the instantaneous kinematics and therefore do not account for any history-dependent effects. Exceptions are the possible inclusion of the Wagner effect~\citep{Wagner_1925} or other time-history formulations~\citep{Reijtenbagh_2026}. In the present work, history-dependent contributions are deliberately omitted to distinguish the entire contribution of kinematic time-history effects from the force depending on instantaneous kinematic state. This totality of time-history effects on force generation, we refer here to as the wake interaction force. Consequently, this wake interaction force can be estimated by comparing the total force acting on the plate with the force predicted by a QS model.

For the present kinematics, the QS model reduces to the Morison equation~\citep{Morison_1950}, containing only a translational drag term proportional to $\vp^2$ and an added mass term proportional to $\accp$. All other QS contributions vanish due to the absence of rotational motion. The associated coefficients are obtained by regression from the numerically-simulated force data: the added mass coefficient is determined from the initial force peak during acceleration, when the plate velocity remains small, while the translational force coefficient is obtained from the asymptotic drag during constant-velocity translation. This procedure yields an added mass coefficient of approximately $11.19$ and a drag coefficient of $1.85$. The obtained added mass agrees well with the theoretical prediction for a finite flat plate, given by $m=K\pi\rho b a^2/4$, where $a$ is the chord length, $b$ the span, $\rho$ the fluid density, and $K$ a coefficient depending on the aspect ratio~\citep{Patton_1965}. This expression yields an added mass coefficient of $11.78$ for the present geometry. The drag coefficient for a flat plate at the much higher Reynolds number $\Rey=10^5$, is reported to be approximately $1.12$~\citep{Blevins_2003}. 

\section{Results}
\label{sec:Results}

\subsection{Flow Observations}
\label{subsec:flow_obs}

\begin{figure}
    \centering
    \includegraphics[width=0.25\linewidth]{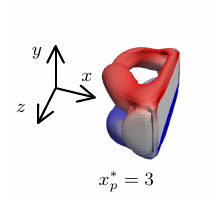}%
    \includegraphics[width=0.25\linewidth]{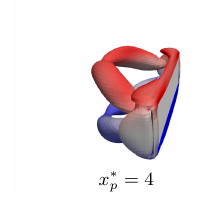}%
    \includegraphics[width=0.25\linewidth]{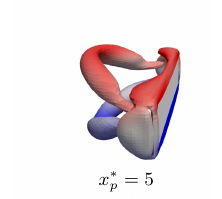}%
    \includegraphics[width=0.25\linewidth]{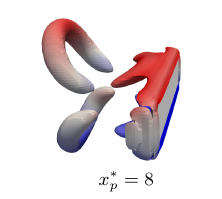}%
    \hfill
    \includegraphics[width=\linewidth]{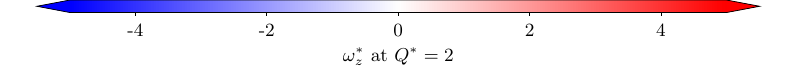}
    \caption{Isosurfaces of $Q$-criterion ($\Qstar=2$), shaded by the $z$-direction vorticity, $\omegazstar$, at selected phases during forward motion for $\Rey=100$.}
    \label{fig:Q_Re100}
\end{figure}

\begin{figure}
    \centering
    \includegraphics[width=0.25\linewidth]{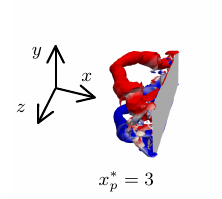}%
    \includegraphics[width=0.25\linewidth]{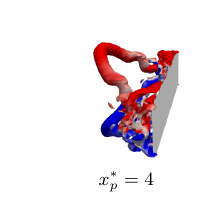}%
    \includegraphics[width=0.25\linewidth]{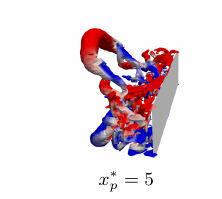}%
    \includegraphics[width=0.25\linewidth]{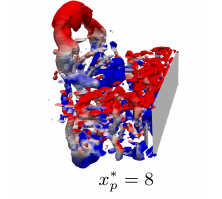}%
    \hfill
    \includegraphics[width=\linewidth]{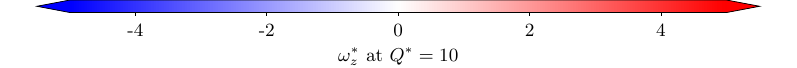}%
    \caption{Isosurfaces of $Q$-criterion ($\Qstar=10$), shaded by the $z$-direction vorticity, $\omegazstar$, at selected phases during forward motion for $\Rey=1000$.}
    \label{fig:Q_Re1000}
\end{figure}

We begin by characterizing the structure of the flow produced by the moving plate. Any nondimensionalisation is performed using length and velocity scales, $c$ and $\vp$, respectively. Isosurfaces of the nondimensional $Q$-criterion ($\Qstar=2$), coloured by the spanwise ($z$-direction) nondimensional vorticity, $\omegazstar$, are shown in figure~\cref{fig:Q_Re100} for $\Rey=100$ at various moments, at which $\xpstar=[3,4,5,8]$, during the plate's forward motion. These illustrate the topology of the three-dimensional vortex structures within the wake. Clear vortex rings can be identified, which grow in time. Though these vortex rings initially reflect the planar, rectangular geometry of the plate, self-induction progressively deforms them into a more rounded, out-of-plane three-dimensional structure~\citep{Straccia_2020}. The wake during forward motion at $\Rey=1000$ exhibits overall similar behaviour (\cref{fig:Q_Re1000}, isosurfaces at $\Qstar=10$). Rather than forming a single coherent vortex ring, additional smaller vortical structures emerge as well. 

\begin{figure}
    \centering 
    \includegraphics[width=\linewidth]{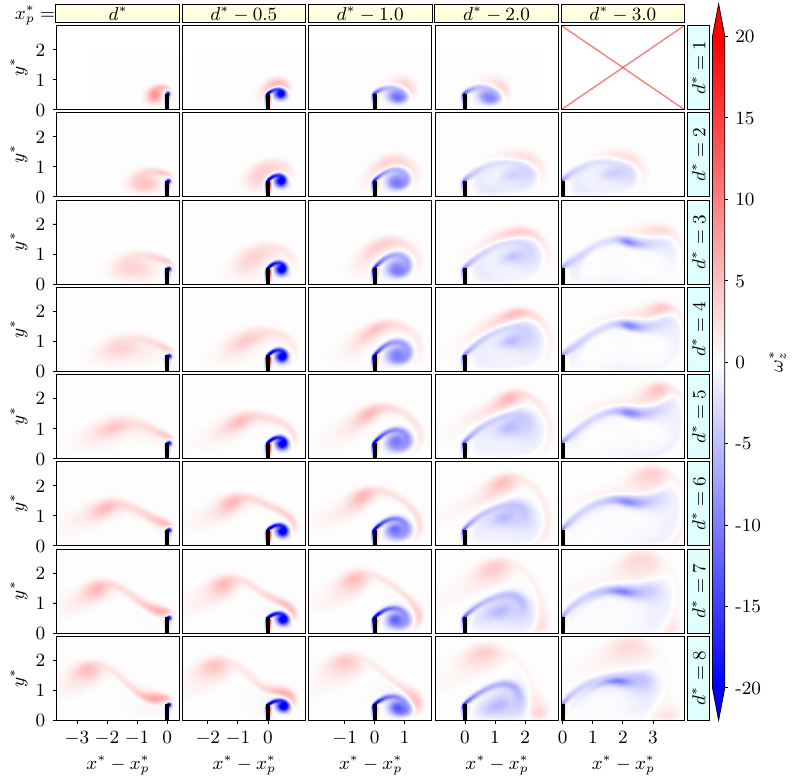} 
    \caption{Spanwise-component, $\omegazstar$, vorticity fields at $\Rey=100$ in the plate $xy$-midplane at selected phases $\dstar$, $\dstar-0.5$, $\dstar-1.0$, $\dstar-2.0$, and $\dstar-3.0$ throughout the post-reversal motion and for various pre-reversal distance cases, $\dstar\in[1,8]$. The top half of symmetric fields are shown, for brevity. A panel containing a red cross indicates data which were not recorded.}
    \label{fig:Re100_fields_xy}
\end{figure}

\begin{figure}
    \centering
    \includegraphics[width=\linewidth]{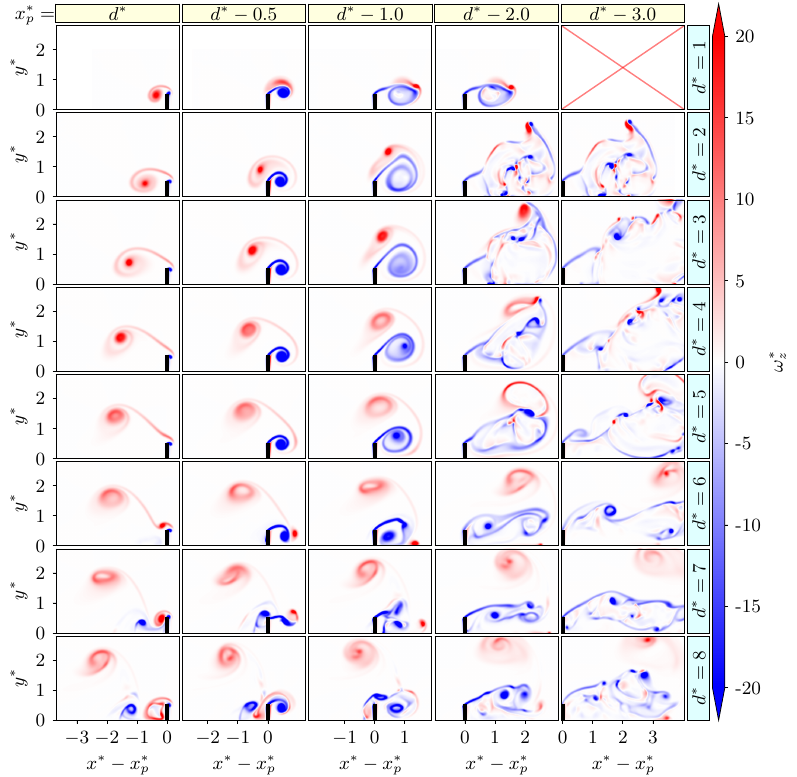}
    \caption{Spanwise-component, $\omegazstar$, vorticity fields at $\Rey=1000$ in the plate $xy$-midplane at selected phases $\dstar$, $\dstar-0.5$, $\dstar-1.0$, $\dstar-2.0$, and $\dstar-3.0$ throughout the post-reversal motion and for various pre-reversal distance cases, $\dstar\in[1,8]$. The top half of symmetric fields are shown, for brevity. A panel containing a red cross indicates data which were not recorded.}
    \label{fig:Re1000_fields_xy}
\end{figure}

Given the complex structure of the wake, comparison across the various cases, both before and after plate reversal, is performed via examination of the flow in the $xy$-aligned midplane. Fields of the nondimensional $z$-component vorticity, $\omegazstar$, in this plane are shown in~\cref{fig:Re100_fields_xy} ($\Rey=100$) and~\cref{fig:Re1000_fields_xy} ($\Rey=1000$). Videos of these vorticity fields are provided for several example cases as Supplementary Material. The flow field is symmetric about the plane $(x,z)=(0,0)$, so for compactness only the upper half of the domain is plotted. Each row corresponds to a specific value of $\dstar$ (indicated at right of figure, shortest pre-reversal displacement at upper row, with larger displacements towards the bottom), while columns represent successive phases of the motion: the first column depicts the instant of reversal, the second column corresponds to a moment at which the plate displacement post-reversal is half a chord length, and subsequent columns show later stages of the reversal motion (indicated at top of figure).

Considering the vorticity fields at $\Rey=100$, presented in~\cref{fig:Re100_fields_xy}, we see that the wake topology at the moment of reversal (left-most column) depends strongly on $\dstar$. For low $\dstar$, the wake consists primarily of a diffuse vortex structure located near the plate. At higher $\dstar$, this vortex structure extends further downstream of the plate and increasingly resembles a diffuse shear layer. This shear layer is, however, non-uniform in strength and width, exhibiting regions of concentrated vorticity associated with both the original vortex core at its most downstream extremity, and also in the vicinity of the plate edge. Upon reversal, the wake structures continue to carry forward momentum, and so do not track with the plate's reversal. Instead, as the plate translates in the negative $x$-direction, the vorticity formed during the forward motion continues to move in the positive $x$-direction. It is displaced transversely in $y$ by the passage of the plate, and is stretched around the plate's boundary. This stretching is more pronounced in lower $\dstar$ cases, where the forward-motion starting vortex and shear layer have remained compact and close to the centreline ($\ystar=0$) by the moment of reversal, but is less severe in higher $\dstar$ cases where, by the moment of reversal, the starting vortex has already undergone significant lateral displacement in $y$, and therefore interacts less strongly with the plate. The presence of the initial starting vortex influences the formation and evolution of the opposite-sign starting vortex during reversal: the reverse starting vortex's roll-up appears inhibited, such that it forms a less compact structure with a larger spatial extent. Moreover, the state of development of the starting wake, which varies with $\dstar$, has a pronounced effect. For low $\dstar$, the small starting vortex largely follows the newly-forming reverse starting vortex, and comes to be located on the leeward side of the reversing plate. For moderate $\dstar$ values, the starting vortex and reverse starting vortex form a pair, which travels outward in $y$ while advecting towards positive $x$. 
At the upper end of the range of $\dstar$ values tested, the starting vortex has already, at the moment of reversal, advected downstream and the initial interaction of the newly-forming reverse starting vortex is with the shear-layer. The  result is that advection of the vortex pair in the positive $x$ direction is less pronounced, and the newly-forming reverse wake remains more compact. 

At $\Rey=1000$, the forward-motion wake topology and newly-forming wake during reversal exhibit superficial similarities to those at $\Rey=100$: A starting vortex is observed to form during forward motion, whose $x$-location relative to the plate increases with greater $\dstar$, and during reversal a reverse-starting vortex is formed, which interacts with the initial starting vortex, forming a vortex pair whose trajectory depends on the value of $\dstar$. These behaviours are illustrated by the $xy$-midplane $\omegazstar$ vorticity fields shown in~\cref{fig:Re1000_fields_xy}. At this higher Reynolds number however, the vortex structures remain more compactly distributed, as a result of the relatively reduced influence of viscous diffusion. Additionally, at sufficiently large values of $\dstar$, the shear-layer connecting the initial starting vortex to the edge of the plate is disrupted and the starting vortex ring undergoes pinch-off after which it appears to develop a hollow, irrotational core. Subsequently, a secondary, compact vortex ring is formed. The accumulation of elevated vorticity observed in the starting shear-layer close to the plate edge at $\Rey=100$ and high $\dstar$ suggests the onset of a similar pinch-off process; however, the actual pinch-off and subsequent vortex formation appear to be suppressed within this range of $\dstar$ by viscous diffusion. At $\Rey=1000$, $\dstar \gtrsim 7$, additional vorticity-bearing structures are observed in the near-wake of the plate at the moment of reversal, namely a small negative-signed vortex on the leeward side of the plate. Upon reversal, the interaction between the plate and the vortex ring appears to fragment the coherent vortex structure into numerous small-scale vortical structures. The resulting wake is substantially more complex than in the lower Reynolds number case. These observations demonstrate that wake interaction is influenced by both the Reynolds number and, more prominently, $\dstar$.

\subsection{Drag Amplification During Reversal}

\begin{figure}
    \centering
    \hfill
    \subcaptionbox{\centering\hspace{-1cm}\label{fig:cd_QS_ds4}}{\includegraphics[width=0.5\textwidth]{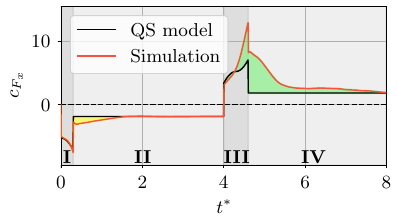}}%
    \subcaptionbox{\centering\hspace{-1cm}\label{fig:veff_ds4}}{\includegraphics[width=0.5\textwidth]{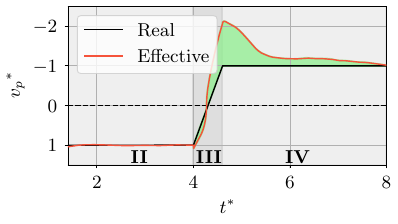}}%
    \hfill\null
    \caption{(a) Force coefficient $\Cfx$ in the $x$-direction over time, for a case with $\dstar=4.0$ at $\Rey=1000$. Force  coefficients computed via numerical simulation are given by the red curve, while a prediction using a QS model is given in black. (b) Real plate velocity and effective velocity. Discrepancies between simulation and QS prediction during forward and reverse motion shaded in yellow and green, respectively. Roman numerals delineate distinct phases of the motion history: \textbf{I}. linear acceleration, \textbf{II}. translation at constant forward velocity, $\vpstar$, in $\xstar$, \textbf{III}. linear deceleration, \textbf{IV}. translation at constant reverse velocity, $-\vpstar$.}
    \label{fig:cd_QS_veff_ds4}
\end{figure}

Given the observed interaction of the plate with its wake during reversal, and the qualitative variation of the form of this interaction with $\dstar$ and $\Rey$, we expect there to also be an effect on the fluid-dynamic forces experienced by the plate.  
For the example kinematic case of $\dstar=4.0$, $\Rey=1000$, the $x$-direction force, $F_x$, nondimensionalised using the fluid density, plate area, and maximum plate velocity as $\Cfx=\Fx/(\nicefrac{1}{2} \rho cb\vpsquare)$, is presented in~\cref{fig:cd_QS_ds4}, together with a QS estimate derived from identical kinematics. 

The QS model predicts a negative force peak during the initial forward acceleration, followed by a constant negative force during the subsequent constant-velocity phase. During the reversal acceleration, it predicts a positive force peak, after which the force returns to a constant positive value during the constant reverse velocity phase. During the forward motion, the numerical result agrees closely with the QS prediction, except immediately after the acceleration, where the simulation exhibits a small start-up effect not captured by the QS model (yellow shaded area). In contrast, during reverse motion, the QS estimate substantially underpredicts the drag (drag during reversal being in the positive $x$-direction), both at the end of the acceleration and during the constant velocity phase (green shaded area). The only distinction between the forward and reverse motion is the presence of wake structures during the reverse phase, concluding that the observed difference in drag arises from the interaction of the plate with these structures. We refer to the discrepancy between the QS prediction and computed drag during reversal henceforth as the wake interaction force, given nondimensionally as $\cdwake = \cFxSim - \cFxQS$.

A conceptually-simple, first-order explanation of the elevated drag during reversal is to attribute it to an effective velocity component induced by the wake through which the plate traverses. This wake-induced velocity alters the relative velocity of the plate with respect to the fluid, compared with the case of translation through quiescent fluid. Assuming that drag scales with the square of the relative velocity, we can compute a single `effective' velocity, $\vpeffstar$, that captures the average effect of the spatially varying wake. This quantity is shown for the reversal phase in~\cref{fig:veff_ds4}, and compared to the actual plate velocity, $\vpstar$, with the difference shaded in green. It can be seen that $\vpeffstar$ briefly exceeds twice $\vpstar$, where the factor $2$ arises from the plate reversing at equal but opposite velocity to its forward motion, indicating that the mean velocity in the wake just behind the plate is similar to the plate's velocity. If the drag force during constant velocity motion scales with the square of the velocity, it is expected to briefly exceed $4$ times the steady state QS-predicted value, as can indeed be seen immediately post-deceleration in~\cref{fig:cd_QS_ds4}. 

\subsection{Temporal Scaling Behaviour of Wake-Induced Drag} 
\label{subsec:cd_wake_t}

\begin{figure}
    \subcaptionbox{\centering\label{fig:cd_wake_all}}{\includegraphics[width=0.65\textwidth]{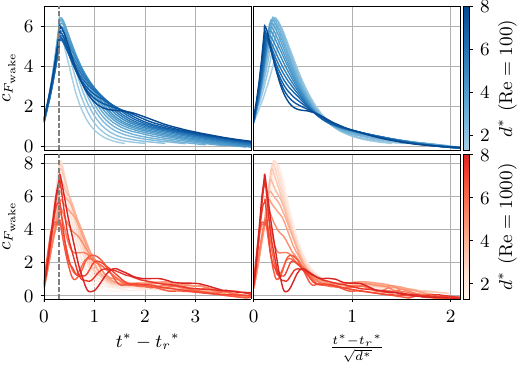}}%
    \hfill
    \subcaptionbox{\centering\hspace{-1.5cm}\label{fig:cd_wake_peak_timing}}{\includegraphics[width=0.28\textwidth]{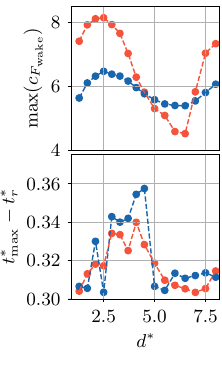}}%
    \hfill\null
    \caption{(a) Wake interaction force, $\cdwake$, over time since reversal (left) and over post-reversal time scaled by $\sqrt{\dstar}$ (right). The dashed vertical line indicates the moment at which deceleration stops. (b) Magnitude, $\mathrm{max}(\cdwake)$, (upper panel) and timing, $\tmaxstar$, (lower panel) of the $\cdwake$ post-reversal peak.}
\end{figure}

The wake interaction force coefficient, $\cdwake$, is obtained for each individual $\dstar$ case by subtracting the QS estimate from the numerically-computed drag. The plate reverses at different times $\trstar=\tstar(\xpstar=\dstar)$ in each case, so $\cdwake$ is plotted against nondimensional time, $\tstar - \trstar$, post-plate-reversal. This allows direct comparison across the various $\dstar$ cases in~\cref{fig:cd_wake_all}~(left), for each Reynolds number.

A key feature is the initial peak in $\cdwake$, whose timing and magnitude vary across cases. To quantify these variations,~\cref{fig:cd_wake_peak_timing} shows the peak magnitude together with its occurrence time post reversal, $\tmaxstar$. Only peaks occurring after the acceleration phase has ended ($\tstar \gtrsim 0.3$) are considered. For intermediate values of $\dstar$, the peak occurs slightly later, likely because the vortex ring is located farther from the plate at the onset of reversal. At larger $\dstar$, vortex pinch-off and the subsequent formation of the secondary vortex ring shift the peak back to earlier times, as the peak interaction is then caused by the secondary vortex ring. Despite these differences, the peak occurs at nearly the same time after reversal for all cases, typically shortly after the acceleration phase ends. This suggests that the wake immediately behind the plate is the primary contributor to force production during reversal. In contrast, the peak magnitude varies more significantly: it initially increases with $\dstar$ before decreasing and rising again.

Returning to~\cref{fig:cd_wake_all}~(left), in addition to the initial primary peak in $\cdwake$, a secondary smaller peak is observed at a later post-reversal time in the highest $\dstar$ cases at $\Rey=1000$, varying both in timing and magnitude with $\dstar$. This, together with the non-monotonic variation in primary peak magnitude with $\dstar$, is hypothesised to be linked to the vortex formation, pinch-off, and subsequent vortex formation discussed in~\cref{subsec:flow_obs,sec:opt_vort_form}. 

Following the primary peak in $\cdwake$, and the secondary peak where present, $\cdwake$ gradually decreases. The rate of this decrease depends strongly on $\dstar$, with higher values of $\dstar$ resulting in a slower decay. This reflects the longer development time during forward motion, and the therefore spatially longer wake at higher $\dstar$. The timescale over which the plate, during its reverse motion, traverses its wake is connected to the spatial length scale of the wake at the moment of reversal, thus modulating the timescale of decay of $\cdwake$. As the decay timescale increases with $\dstar$, it follows naturally to seek a scaling for the temporal axis with a factor dependent on $\dstar$. Using $\sqrt{\dstar}$ for this factor produces a reasonable collapse of the late-time decay, as shown in~\cref{fig:cd_wake_all}~(right). The successful collapse suggests that the observed variation in decay rate can largely be captured by a single $\dstar$-dependent timescale set by the state of the wake at the moment of reversal. To investigate this hypothesis, the development of the streamwise wake velocity during forward motion is examined in the following section.

\subsection{Wake Velocity as a Proxy for Interaction Force}
\label{subsec:U_w_t}

\begin{figure}
    \centering
    \includegraphics[width=\linewidth]{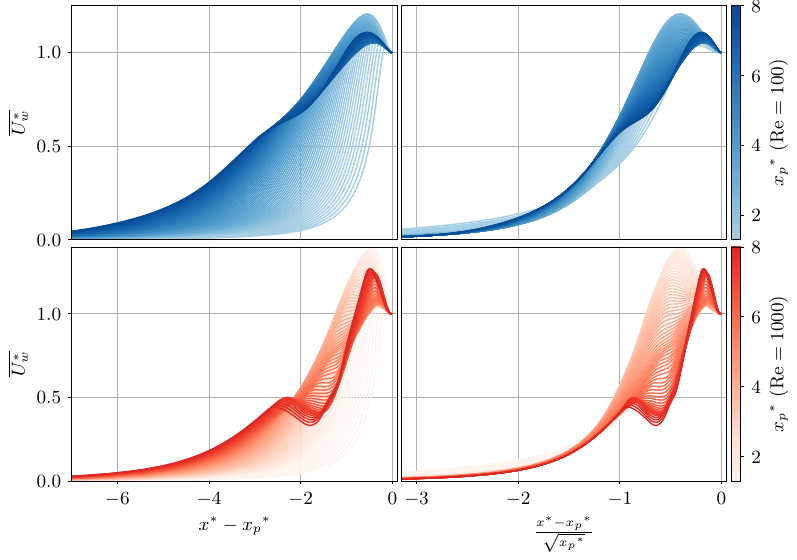}
    \caption{Mean streamwise wake velocity for various nondimensional distances traversed, $\xpstar$. Plotted against relative distance behind the plate (left) and against scaled distance behind the plate (right). Data shown for $\Rey=100$ (blue) and $\Rey=1000$ (red).}
    \label{fig:U_wake_all}
\end{figure}

To find an argument explaining the observed scaling in the decay of $\cdwake$, we examine the development of the streamwise velocity in the wake, $\Uwnoavg$. It offers an estimate of the flow that the plate encounters during reversal. This relates to the already-introduced concept of effective velocity, $\vpeff$ (\cref{fig:veff_ds4}), which can be interpreted as the sum of $\vp$ and a wake-imposed component, motivating a focused examination of $\Uwnoavg$. To obtain a representative value of the wake velocity, the nondimensional velocity field is averaged over the projected area of the plate ($-c/2 \le y \le c/2,\; -b/2 \le z \le b/2$) to obtain $\Uwstar$. This wake velocity, measured as a function of the distance behind the plate, $\xstar - \xpstar$, can be computed at various stages of forward motion (after translating various distances $\xpstar$), capturing the evolution of the wake. The computed values of $\Uwstar$ are given for both Reynolds numbers in~\cref{fig:U_wake_all}~(left). 

The spatial development of $\Uwstar$ with increasing nondimensional distance travelled, $\xpstar$, closely mirrors the behaviour of $\cdwake$ over nondimensional time. Firstly, irrespective of $\xpstar$, there is a peak at a similar, low $\xstar - \xpstar$ with a magnitude first increasing, then decreasing with $\xpstar$. Furthermore, a second peak occurs for the $\Rey=1000$ cases at large $\xpstar$. Lastly, for higher $\xpstar$, $\Uwstar$ remains high over a longer length scale, similar to how $\cdwake$ decays slower in time after reversal when $\dstar$ is high. Similar to the $\sqrt{\dstar}$ scaling applied to the temporal axis in the $\cdwake$ evolution in~\cref{fig:cd_wake_all}, an analogous $\sqrt{\xpstar}$ scaling is applied to the downstream distance behind the plate, $\xstar-\xpstar$. This results in a substantial collapse of the wake velocity profiles, particularly for $\Rey=1000$, as shown in~\cref{fig:U_wake_all}~(right). 

This collapse reflects that the wake grows in time, meaning its spatial decay occurs further downstream for higher $\xpstar$. It is evident that the observed scaling in $\cdwake$ is thus rooted in a similar scaling in $\Uwstar$, since the plate must traverse this excess velocity field when reversing. The remaining question is: what physical mechanism underlies the downstream scaling of the wake velocity?

\subsection{Vortex Dynamics and Scaling Origin} 
\label{subsec:vortex_dym}

\begin{figure}
    \subcaptionbox{\centering\label{fig:tracking_schematic}}{\includegraphics[width=0.43\textwidth]{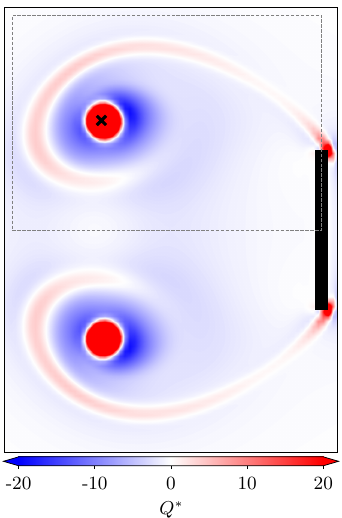}}%
    \hfill
    \subcaptionbox{\centering\hspace{-1.45cm}\label{fig:tracked_xy_circulation}}{\includegraphics[width=0.5\textwidth]{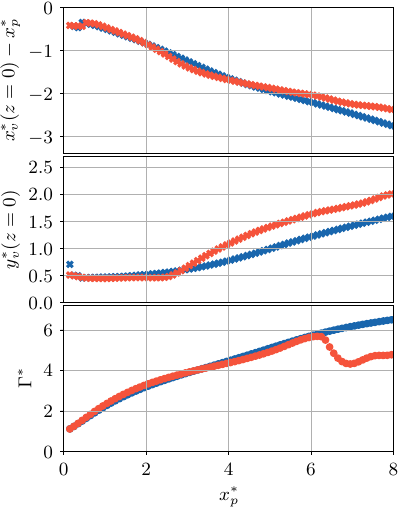}}
    \caption{(a) Example spatial distribution of the nondimensional $Q$-criterion, $\Qstar$, in the $xy$-midplane at $\Rey=1000$ after the plate has travelled $\xpstar=3$, illustrating the tracked vortex-core location (black cross marker). The region used for determining circulation via an area integral of the vorticity field is demarcated by a grey, dashed line. (b) Tracked vortex location in the $xy$-midplane ($z=0$) (upper two panels)  and circulation (lower panel)  over distance, $\xpstar$, travelled by the plate during forward motion. Data shown for $\Rey=100$ (blue markers) and $\Rey=1000$ (red markers).}
    \label{fig:vortex_tracking_all}
\end{figure}

\begin{figure}
    \centering
    \subcaptionbox{\centering\label{fig:vortex_tracking_3D_Re100}}{\includegraphics[width=0.5\textwidth]{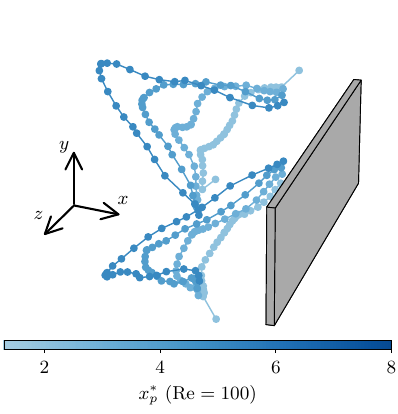}}%
    \subcaptionbox{\centering\label{fig:vortex_tracking_3D_Re1000}}{\includegraphics[width=0.5\textwidth]{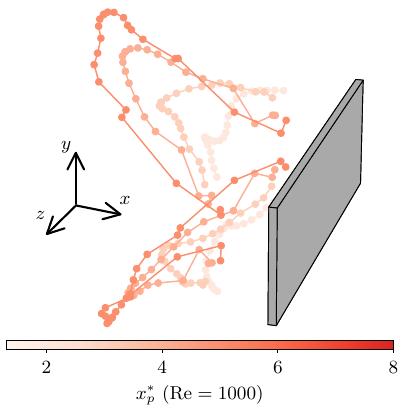}}%
    \hfill
    \caption{Vortex ring coordinates from three-dimensional tracking at $\xpstar=2$, $3$, $4$, and $5$ (in order of increasing colour saturation). (a) $\Rey=100$ (blue) and (b) $\Rey=1000$ (red).}
    \label{fig:vortex_tracking_3D}
\end{figure}

We seek to uncover the origin of the observed scaling of $\Uwstar$. We therefore examine the underlying vortex dynamics during the formation of the wake. Vortex core locations are identified by computing the centroid of regions of positive $Q$-criterion, weighted by their $Q$ value. The starting vortex ring core's intersection with the $xy$-midplane is found using this definition, as illustrated in~\cref{fig:tracking_schematic}. Repeating this process over the forward motion yields~\cref{fig:tracked_xy_circulation}~(top two panels). The $xy$-midplane vortex' nondimensional downstream distance from the plate increases roughly linearly with $-0.32$ per unit formation distance $\xpstar$. This means the vortex moves with the plate, albeit slightly slower. The transverse position remains nearly constant initially, after which it deviates roughly linearly from $\xpstar \approx 2.5$ onwards with slope $0.30$ per unit formation distance $\xpstar$. This lateral drift in the $xy$-midplane reflects the reshaping of the vortex ring, more evident when visualising the starting-vortex core locations in three dimensions~(\cref{fig:Q_Re100,fig:Q_Re1000}). A measure of the vortex strength, its circulation, can be estimated by integrating the vorticity within a rectangular region behind the plate shown in~\cref{fig:tracking_schematic}, resulting in~\cref{fig:tracked_xy_circulation}~(bottom). As circulation is conserved along a vortex ring, this is a representative value for the ring's actual circulation. This circulation initially increases rapidly after which its increase slows down. Both the vortex trajectories and circulation only mildly depend on $\Rey$. 

To estimate the three-dimensional starting-vortex ring shape, the vortex core is identified separately at various spanwise ($xy$) planes. The resulting topology is depicted at various time instants throughout the forward motion in~\cref{fig:vortex_tracking_3D}. This reveals that, rather than simply tending towards a circular ring shape over time, the initially elongated ($AR\approx 4$) ring also deforms in the streamwise direction, with the vortex core near the plate ends remaining in the vicinity of the plate, while near the $xy$-midplane the vortex core advects significantly downstream, as quantified in~\cref{fig:tracked_xy_circulation}~(top). 

\begin{figure}
    \centering
    \includegraphics[width=\linewidth]{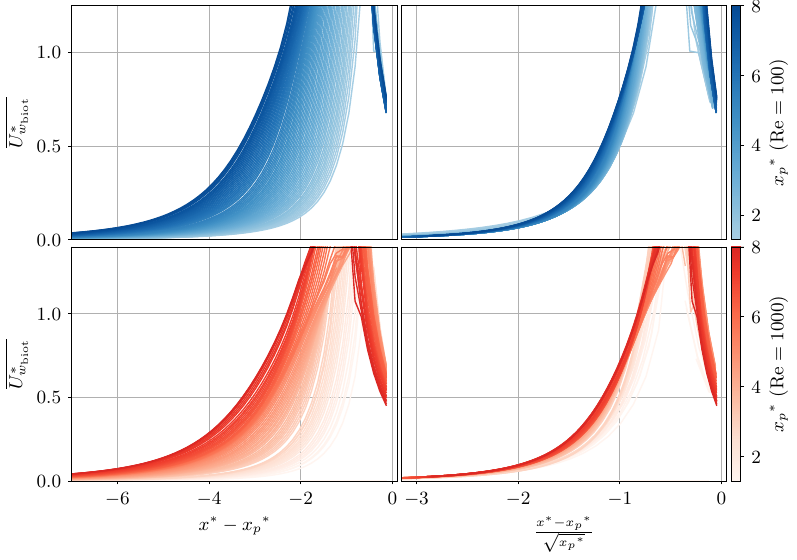}
    \caption{Mean streamwise wake velocity, $\Uwbiotstar$, during plate forward motion, computed by applying the Biot--Savart law to the starting vortex ring, assuming concentrated circulation at the vortex-ring centreline. Each curve corresponds to a nondimensional distance traversed, $\xpstar$ ($\xpstar$ increasing with increasing colour saturation). Data are plotted against relative distance behind the plate (left) and against scaled distance behind the plate (right). Data shown for $\Rey=100$ (blue) and $\Rey=1000$ (red).}
    \label{fig:U_wake_biot_all}
\end{figure}

The induced wake velocity is estimated by discretising the tracked three-dimensional starting-vortex ring into individual vortex filaments and applying the Biot--Savart law to each filament, assuming concentrated vorticity along the vortex centreline. The streamwise component, $\Uwbiotnoavg$, is nondimensionalised and averaged over the projected area behind the plate as before to obtain $\Uwbiotstar$. As was done for $\Uwstar$ in~\cref{fig:U_wake_all}, $\Uwbiotstar$ is plotted over the distance behind the plate for various moments during the forward motion in~\cref{fig:U_wake_biot_all}~(left), for both $\Rey=100$ and $\Rey=1000$. This shows the evolution of the flow velocity in the wake, as induced by the starting vortex ring modelled as a line vortex at its core. Although considering only a single vortex ring, neglecting viscous effects, and excluding the plate, represents a simplified model of the wake dynamics, it provides a useful means of identifying the processes that significantly influence the wake flow and its decay. Interestingly, this approach reproduces qualitatively similar trends to the actual $\Uwstar$ (\cref{fig:U_wake_all}~(left)), particularly that the velocity decay occurs progressively farther downstream with increasing $\xpstar$. The remaining differences arise from the simplifying assumptions: the peak velocity is overestimated and, for the $\Rey=1000$ case, no secondary peak is present because the model contains only a single vortex ring. Applying the same spatial scaling as in~\cref{fig:U_wake_all}~(right) yields~\cref{fig:U_wake_biot_all}~(right), which shows a qualitatively similar collapse of the decay. We yield from this approach the key insight that the wake evolution and its far-field decay scaling are governed by the vortex dynamics, in particular the vortex location, shape, and strength. 

\subsection{Simplified Two-Dimensional Vortex}
\begin{figure}
    \centering
    \hfill
     \subcaptionbox{\centering\label{fig:schematic_point_vortex}}{\includegraphics[width=0.43\textwidth]{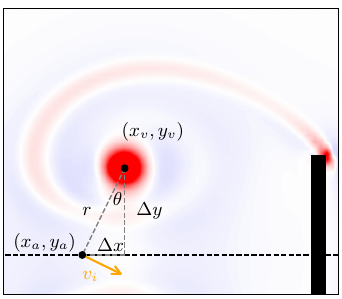}}%
    \hfill
    \subcaptionbox{\centering\hspace{-1.25cm}\label{fig:U_wake_decay_comparison}}{\includegraphics[width=0.5\textwidth]{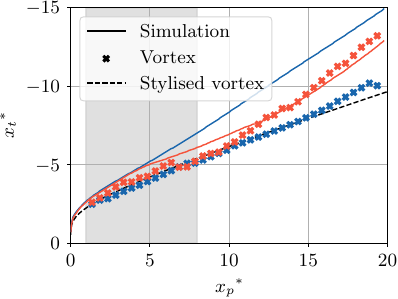}}%
    \null
    \caption{(a) Schematic indicating induced velocity at point $A=(\xa, \ya)$ along the centreline by a point vortex at point $V=(\xv, \yv)$. (b) Lengthscale $\xtstar$ over which the streamwise nondimensional velocity in the wake during forward plate motion exceeds threshold $\alphastar$ (chosen here as $\alphastar = 0.05$). Values derived from: actual numerically-simulated velocity (solid curves), a point vortex with observed circulation and averaged position (markers), and a stylised point vortex (dashed curve). $\Rey=100$ (blue) and $\Rey=1000$ (red) shown. The gray shaded region indicates the range of $\dstar$ investigated in the present study.}
\end{figure}

So far, we have observed the temporal decay of $\cdwake$ and the spatial decay of $\Uwstar$ to exhibit similar behaviour, scaling proportionally to $\sqrt{\dstar\vphantom{\xpstar}}$ and $\sqrt{\xpstar}$, respectively. The previous section suggested that this scaling is rooted in the vortex dynamics. Here, we investigate why the observed combination of vortex location and strength produces the apparent square-root scaling. For this purpose, we model the decay of the streamwise wake velocity during the forward phase of the motion up to $\xpstar=20$ using a simplified two-dimensional point vortex of circulation $\Gamma$, thereby capturing the long-term wake formation process. A schematic of the two-dimensional point-vortex model is shown in~\cref{fig:schematic_point_vortex}. The $x$-component velocity at a generic point $A=(\xa, \ya)$, induced by a point vortex at $(\xv, \yv)$, is $v_{i_x} = \frac{\Gamma cos(\theta)}{2\pi r}$, where $r$ represents the distance between the vortex and point $A$. Using $cos(\theta)=\frac{\Delta y}{r}$ and $r^2 = \Delta x^2 + \Delta y^2$, we can rewrite this expression as $v_{i_x} = \frac{\Gamma \Delta y}{2\pi (\Delta x^2 + \Delta y^2)}$. Rearranging this expression for $\Delta x$ gives $\Delta x^2=\frac{\Gamma \Delta y}{2\pi v_{i_x}} - \Delta y^2$.

To quantify the length of the wake, we consider a coordinate behind the plate $x_t$ along the centreline for which the induced velocity reduces below a certain threshold $v_{i_x}=\alpha$. This threshold is met a certain distance $\Delta x$ behind $\xv$, meaning the length scale can be expressed as $\xt=\xv+\Delta x$. The previous expression of $\Delta x^2$ can be simplified with $v_{i_x}=\alpha$ and $\Delta y=y_v$, as we take the coordinate along the centreline ($y_a=0$), resulting in $\Delta x = -\sqrt{y_v(\frac{\Gamma}{2\pi \alpha} - y_v)}$. Thus, the wake lengthscale can be expressed in nondimensional form as 
\begin{equation}
\label{eqn:xtstar}
\phantom{.}\xtstar = \xvstar + {\Delta x}^* = \xvstar - \sqrt{\yvstar(\frac{\Gammastar}{2\pi \alphastar} - \yvstar)}.
\end{equation}

\begin{figure}
    \subcaptionbox{\centering\hspace{-1.25cm}\label{fig:vortex_xpath_mean}}{\includegraphics[width=0.465\textwidth]{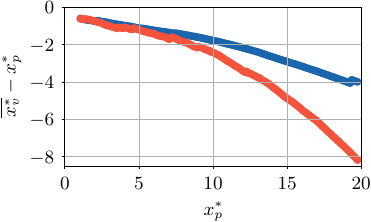}}%
    \hfill
    \subcaptionbox{\centering\hspace{-1.25cm}\label{fig:vortex_ypath_mean}}{\includegraphics[width=0.465\textwidth]{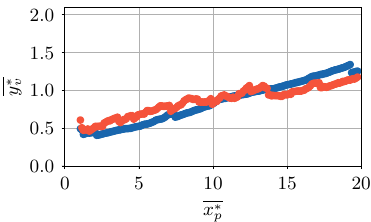}}%
    \caption{Tracked spanwise-averaged vortex-center location. (a) Coordinate, $\xvstarbar$, in the direction of plate motion, and (b) coordinate, $\yvstarbar$, in the transverse direction, as a function of the plate position, $\xpstar$, during forward motion. Data are shown for $\Rey=100$ (blue markers) and $\Rey=1000$ (red markers).}
    \label{fig:vortex_paths_mean}
\end{figure}

This relation expresses an estimate for the nondimensional wake length in terms of only the nondimensional circulation, $\Gammastar$, of the two-dimensional point-vortex, and its nondimensional coordinates, $(\xvstar, \yvstar)$. To be able to apply this simplification to the simulated cases, a representative two-dimensional vortex location is found based on the three-dimensional vortex rings in~\cref{fig:vortex_tracking_3D} by averaging over half of the symmetric vortex ring in the spanwise direction. This yields spanwise-averaged coordinates $(\xvbar, \yvbar)$ at each plate position, $\xpstar$, as shown in~\cref{fig:vortex_paths_mean}. Note that the behaviour of these curves is distinct from that of the vortex cores at the $xy$-centreplane~(\cref{fig:tracked_xy_circulation}). While the vortex-ring intersection with the $xy$-midplane translates substantially in both the $x$- and $y$-directions, the spanwise-averaged vortex-ring position, $(\xvstarbar, \yvstarbar)$, initially moves only slowly and approximately linearly in the streamwise direction, $x$, with negligible lateral displacement in $y$. After $\xpstar \approx 10$, the streamwise translation remains approximately linear, but with a steeper slope for both $\Rey=100$ and $\Rey=1000$. Initially, the nondimensional vortex position changes at rates of approximately $-0.15$ and $0.045$ in the $x$- and $y$-directions, respectively, per unit formation distance $\xpstar$. After $\xpstar=10$, the streamwise translation rate increases most markedly for the $\Rey=1000$ case, reaching approximately $-0.5$ per unit formation distance. We hypothesise that the origin of the increased streamwise advection rate at large $\xpstar$ is that, once fully detached from the plate, the vortex ring is no longer advected by the plate and instead propagates primarily through its own, weaker self-induction.

The nondimensional circulation, $\Gammastar$, of the starting vortex ring, shown earlier in~\cref{fig:tracked_xy_circulation}~(bottom), grows rapidly early-on, while the rate of circulation growth reduces with $\xpstar$. Once the plate reaches constant velocity, the circulation appears to follow an approximately square-root dependence on time. This observed behaviour is consistent with theoretical predictions for an impulsively started plate: according to the general solution to Stokes' first problem, circulation scales as the square root of time~\citep{Reijtenbagh_2023}. Since the acceleration phase of the motion is relatively short, the plate behaves almost impulsively, providing a physical motivation for the existence of a square-root scaling. This circulation, $\Gammastar\sim\sqrt{\xpstar}$, derived via integration of the vorticity in the $xy$-midplane, is used as an estimate of the whole vortex ring circulation, knowing that circulation is conserved along the vortex filament.

From our findings regarding the scaling of the decay of the forces~
(\cref{fig:cd_wake_all}) and of the streamwise wake velocity~(\cref{fig:U_wake_all}), it is expected that the wake length scale $\xtstar$ would increase approximately with $\sqrt{\xpstar}$. The distance over which the streamwise component of the wake velocity in the simulated velocity fields decays to below a threshold of $\alphastar=0.05$ is shown for both tested Reynolds numbers as solid lines in~\cref{fig:U_wake_decay_comparison}. The threshold distance based on the simplified two-dimensional point vortex model using spanwise-averaged vortex locations from~\cref{fig:vortex_paths_mean} and circulation from~\cref{fig:tracked_xy_circulation}~(bottom) is shown using markers in~\cref{fig:U_wake_decay_comparison}. These two trends behave qualitatively similarly to one another. Importantly, in the intermediate range of $\dstar\in[1,8]$ (the area highlighted in grey) for which we have simulated plate reversal motions, the difference between the shape of these curves and a square-root is not easily discernable. Observing the curves' shape for larger $\xpstar$ however, it becomes apparent that a square-root scaling does not well capture the scaling of $\xtstar$ over this broader range. 

To determine why the wake lengthscale behaves in this manner, it is instructive to consider the scaling of both the vortex location and circulation. The vortex $x$- and $y$-locations we can approximate from~\cref{fig:vortex_paths_mean} to be linear in $\xpstar$ and, since $\dstar\equiv\xpstar(\tstar=\trstar)$, it follows that $\xvstar\sim\dstar$ and $\yvstar\sim\dstar$ (at least for small $\dstar$). Similarly, from~\cref{fig:tracked_xy_circulation}, we can approximate the vortex ring circulation as scaling as $\Gammastar\sim\sqrt{\dstar}$. Using the empirically-determined translation rates of $-0.15$ and $0.045$ per formation distance $\xpstar$ for $\xvstar$ and $\yvstar$, and an empirical coefficient of $2.2$ in the scaling for $\Gammastar$, one obtains the black, dashed curve shown in~\cref{fig:U_wake_decay_comparison}. Comparing this with the curves derived from the simulations and from the point vortex model applied without simplifications on the scaling of the vortex location and strength, one sees that these simplified scalings capture the essential physics.

Substituting these scalings into~\cref{eqn:xtstar}, and ignoring the minimal slope of $\yvstar$, reveals that rather than the square-root dependence assumed in~\cref{fig:cd_wake_peak_timing,fig:U_wake_all,fig:U_wake_biot_all}, under the assumptions of our stylised-vortex model the wake lengthscale is actually governed by a fourth root plus linear dependence on $\xpstar$ and thus $\dstar$. 

For $\xpstar<10$, all curves in~\cref{fig:U_wake_decay_comparison} exhibit qualitatively similar behaviour. In this regime, the assumption of an approximately constant downstream advection rate of the starting vortices remains reasonably valid. The initial rapid increase of $\xtstar$ observed in the simulated velocity data (\cref{fig:U_wake_decay_comparison}, solid curves) at low $\xpstar$ is captured by the fourth-root term of the stylised-vortex model (dashed curve), which dominates at very small $\xpstar$. The linear term in the stylised vortex model dominates at large $\xpstar$, capturing the behaviour of the simulated data accurately. The stylised-vortex model's prediction curve is qualitatively accurate across the entirely range of $\xpstar$ computed, although the value of $\xtstar$ is consistently underpredicted. The linear behaviour of the simulated results (solid curves) at large $\xpstar$, especially at $\Rey=100$, suggests that the point-vortex-based model's prediction of a fourth-root plus linear relation describes the underlying physics more accurately than the initially assumed square-root scaling. The slight departure in linearity at $\Rey=1000$ when $\xpstar>10$ is related to the increase in advection rate around this time seen in~\cref{fig:vortex_xpath_mean}, which is considerably more prominent at $\Rey=1000$. This is not captured by the stylised-vortex model, but is observed when $\xtstar$ is predicted using point vortices placed at the actual observed locations, with observed strength, without any stylisation (\cref{fig:U_wake_decay_comparison}, markers). Up until $\xpstar=10$ the stylised model matches the real point vortex model, before the assumptions driving the stylisation break down after this $\xpstar$. At very low $\xpstar$, where the fourth-root term in the model is predicted to dominate, no location of the starting vortex during its initial formation could be robustly identified to place markers in~\cref{fig:U_wake_decay_comparison}. Nevertheless, the fourth-root match between stylised model and directly-measured $\xtstar$ at low $\xpstar$ gives confidence that the wake velocity and its evolution are indeed fundamentally linked with vortex position and strength.

\subsection{Optimal Vortex Formation and the Energetics of Wake Capture}
\label{sec:opt_vort_form}

Starting-vortex dynamics has been identified as centrally linked to the lengthscale of $\Uwstar$ and, commensurately, the timescale of $\cdwake$. 
Pivotal to the behaviour of the starting vortex ring is the concept of optimal vortex formation~\citep{Gharib_1998, Dabiri_2005}, which dictates that a vortex ring grows in strength only up to a maximum nondimensional translation distance, typically recognised (using our present kinematic nomenclature) as $\xpstaropt \approx4$, before detaching from its generating body. This limitation on formation timescale, lengthscale, and therefore strength of a starting vortex has previously been linked to biopropulsion~\citep{Dabiri_2009} and has recently been identified as an important determining factor in the kinematics of biological flapping flight~\citep{Sun_2025}, but the implications for wake capture have not previously been explored, and it is unclear what the qualitative role is for Reynolds numbers in the range relevant to insect flight.

In the present study, the stroke length $\dstar$ governs the form of the wake at the moment of plate reversal, and as such the form and scaling of the elevated drag force experienced by the plate during the reversal. Whether the starting vortex remains attached or pinches off before the plate reverses direction, depends fundamentally on whether $\dstar$ exceeds $\xpstaropt$. For low $\xpstar$, the vortex ring grows in strength with traversal distance, causing an increased peak in $\cdwake$ as $\dstar$ increases. The magnitude of the peak begins to reduce as $\dstar$ approaches the optimal vortex formation limit. When the plate motion exceeds $\xpstaropt$, the primary vortex detaches and advects downstream, with a second vortex ring forming behind it. As the plate reverses through these two distinct vortex structures at $\Rey=1000$, it experiences two separate regions of high velocity with a dip between them~(\cref{fig:U_wake_all}), explaining the  double peak observed in $\cdwake$~(\cref{fig:cd_wake_all}).

At $\Rey=100$, in contrast, the vortex ring never successfully detaches, thus suppressing the secondary vortex formation and preventing the emergence of a double force peak (\cref{fig:cd_wake_all}). This suppressed pinch-off at $\Rey=100$ can be attributed to stronger viscous diffusion balancing the vorticity generation at this Reynolds number, and may have implications for the optimality of biological flapping kinematics across this Reynolds number range.

\begin{figure}
    \centering
    \includegraphics[width=0.5\linewidth]{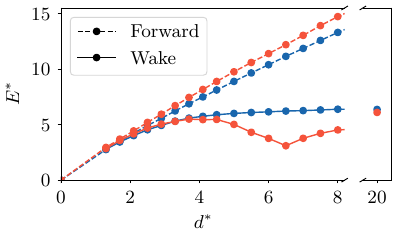}
    \caption{Work, $\Efstar$, done by the fluid on the plate during forward motion to a given $\dstar$ (dashed curve). Work, $\Ewstar$, done by the excess drag force, $\cdwake$, during reverse motion (solid curve) for various pre-reversal translation distances, $\dstar$. Data shown for $\Rey=100$ (blue) and $\Rey=1000$ (red).}
    \label{fig:Energy_from_forces}
\end{figure}

Since optimal vortex formation has been related to the energy contained within the pinching-off vortex ring~\citep{Gharib_1998}, and particularly as wing--wake interaction is often regarded as a mechanism for extracting and converting this energy into fluid forces~\citep{Sane_2003,Lehman_2008}, it is natural to evaluate the energetics of this process. We therefore calculate the nondimensional energy associated with the additional wake interaction force as $\Ewstar=\int{\cdwake \vpstar d\tstar}$. This can be interpreted as the energy extracted from the wake to generate additional drag. $\Ewstar$ is compared to the energy expended due to drag generation during forward motion, $\Efstar$, which is computed similarly, in~\cref{fig:Energy_from_forces}.

The energy expenditure $\Efstar$ during forward motion increases approximately linearly with $\dstar$, as a result of the nearly constant force and constant velocity over most of the forward motion. For low $\xpstar$, $\Ewstar$ is only slightly lower than $\Efstar$, indicating that almost all of the energy expended on the forward stroke is recovered in generating additional drag during the reverse stroke. Beyond a certain $\dstar$ however, $\Ewstar$ no longer increases with $\dstar$ but plateaus. This indicates that for long stroke lengths, the energy transferred into the wake during the forward motion is not efficiently converted into increasing drag during reversal. The plateau value of $\Ewstar$ is reached at $\dstar=\xpstaropt\approx4$, suggesting a link with optimal vortex formation. Indeed, at $\Rey=1000$, $\Ewstar$ reaches a maximum at $\dstar\approx4$ and dips thereafter rather than reaching a simple plateau as at $\Rey=100$, with a local minimum at $\dstar\approx6.5$. A kinematic case with extended translation to $\dstar=20$ before reversal was simulated, revealing that the plateau level of $\Ewstar$ reached at $\dstar\approx4$ remains a limiting value (\cref{fig:Energy_from_forces}, at right).

This observation has implications for the peak value of $\cdwake$. As $\Ewstar$ stagnates with increasing $\dstar$, the integrated area under the $\cdwake$ curve (\cref{fig:cd_wake_all}) no longer increases. Meanwhile, the time scale associated with the decay of $\cdwake$ continues to grow. To maintain a constant integrated $\Ewstar$, the peak value must therefore decrease to compensate for the longer decay. This argumentation aligns with the observations in ~\cref{fig:cd_wake_peak_timing}~(top), up to $\dstar=6.5$, whereafter the peak $\cdwake$ magnitude increases once more. At $\Rey=1000$, the minimum in peak $\cdwake$ magnitude at $\dstar=6.5$ aligns with the local minimum in $\Ewstar$ at the same $\dstar$ value, as the starting vortex ring has shed and the secondary vortex ring has not yet developed strongly. At $\Rey=100$, no local minimum is observed in $\Ewstar$, consistent with the absence of a fully developed pinch-off of the starting vortex ring (\cref{fig:Re100_fields_xy}). Nevertheless, the accumulation of vorticity in the shear layer near the plate edge suggests an incipient pinch-off process, which gives rise to qualitatively similar behaviour in $\mathrm{max}(\cdwake)$ as that observed at $\Rey=1000$.

\section{Conclusion}
\label{sec:Conclusion}

Understanding insect flight fluid mechanics can yield fundamental insights that explain observed behaviours in biological and ecological systems, but is also crucial for advancing applications of unsteady aerodynamics in bio-inspired engineering. A key phenomenon about which we still have limited understanding is wing--wake interaction. Wing--wake interaction is discussed widely, but its origins and scaling are not yet thoroughly understood. This study aimed to begin filling this gap by examining the wing--wake interaction force generated in the abstracted context of a flat plate undergoing a single-degree-of-freedom (DOF) reversing motion. Three-dimensional numerical simulations were performed for this purpose.

A discrepancy between the numerically-computed drag force acting on the plate and that predicted using a quasi-steady model was found during the reverse motion, with drag being higher during reversal. This excess drag, $\cdwake$, was attributed to interaction with the wake structures generated during the forward motion of the plate. The peak value of $\cdwake$, reached shortly after reversal, is on the order of four times the steady-state drag, which was linked to an effective relative velocity between plate and wake flow of approximately double the plate's velocity, and a velocity-squared scaling of the steady-state drag component.

We have associated the temporal dynamics of the excess drag to the location, shape, and circulation, of the starting-vortex ring produced during forward motion, and so to the lengthscale of the wake at the moment of reversal. In so doing, we have refined an initial empirical estimate of $\sqrt{\dstar}$ scaling for the temporal decay of $\cdwake$ and spatial decay of $\Uwstar$, providing a physically-motivated underlying fourth root plus linear scaling dependence on $\dstar$. This dependence matches observations from the numerical simulations well for $\dstar<10$, after which vortex pinch-off during the forward plate affects the qualitative form of the wake. 

The observation of two peaks in $\cdwake$ for high $\dstar$ at $\Rey=1000$ was explained by the detachment of the starting-vortex ring, and subsequent formation of a secondary vortex ring. The appearance of the double-peak wake-induced drag above $\dstar\approx 4$ suggests a relation to optimal vortex formation. An analysis of the total energy retrieved from the wake during reversal to generate excess drag showed that, beyond $\dstar\approx 4$, the energy retrieved from the wake plateaus, providing an upper limit for the efficiency of wake capture for given motion amplitude $\dstar$. In this context of a one-DOF reversing plate, the wake energy retrieval results in only an increased drag, which could be considered disadvantageous in the context of flapping flight, but the dynamics observed here represent an energy availability which, with the correct kinematic strategy, could conceivably be captured for other purposes such as lift or thrust production instead.

\section*{Acknowledgements}
The authors thank M. Lauber for assistance in developing the numerical simulations using \textit{WaterLily.jl}, G.J. Mulder and E. Overmars in helping to prepare the experimental measurement campaign, and J. Westerweel for valuable discussions and the provision of code to compute the effect of the starting vortex ring via the Biot--Savart law. A-J.\ Buchner was supported by the Netherlands Organisation for Scientific Research (NWO), under VENI project number 18176.

\appendix
\renewcommand{\thefigure}{A\arabic{figure}}
\setcounter{figure}{0}

\newpage
\section{Convergence of the numerical simulations}
\label{sec:appendix_convergence}

Studies were performed to test the convergence of the numerical simulations in \textit{WaterLily.jl} with numerical setup parameters. The motion case chosen for these studies was $\dstar=1.0$ with $\Rey=1000$. This case features the smallest tested translation distance and the highest tested Reynolds number, resulting in the smallest relevant length scales. These characteristics make it the most demanding case in terms of spatial discretisation, tested in~\cref{subsec:appendix_discretisation}. The same motion case was used in testing independence on numerical domain extent in~\cref{subsec:appendix_domain}. Finally, the effect of applying a symmetry condition in the numerical domain was checked in~\cref{subsec:appendix_symmetry}.

\subsection{Spatial Resolution}
\label{subsec:appendix_discretisation}
\begin{figure}
    \subcaptionbox{\centering\hspace{-1.2cm}\label{fig:convergence_L}}{\includegraphics[width=0.465\textwidth]{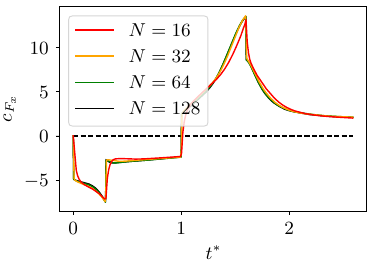}}%
    \hfill
    \subcaptionbox{\centering\hspace{-1.2cm}\label{fig:convergence_L_percentage}}{\includegraphics[width=0.465\textwidth]{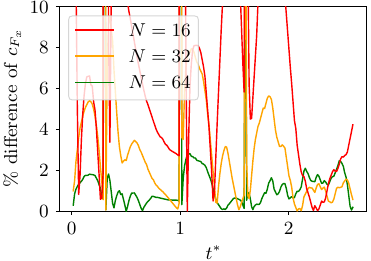}}%
    \hfill\null
    \caption{(a) Nondimensional force component in the $x$-direction, $\Cfx$, for varying discretisation $N$ between $N=16$ and $N=128$ cells across chord length $c$. The performed case has $\Rey=1000$ and $\dstar=1.0$. (b) Difference between the coarser cases and the finest case, expressed as a percentage of the instantaneous force from the finest case.}
    \label{fig:convergence_L_all}
\end{figure}

The numerical grid spatial resolution is varied as $N=16$, $32$, $64$, and $128$ cells per chord length $c$, while using a computational domain for which the solution has converged with respect to domain size. The $x$-component of the numerically-computed force acting on the plate varies only mildly for varying discretisation in this range, especially between $N=64$ and $N=128$~(\cref{fig:convergence_L}). The difference between the lower-resolution cases and the finest-resolution case, as a percentage of finest-case instantaneous values, are shown in~\cref{fig:convergence_L_percentage}. Large momentary differences can be observed near the start and end of accelerations. These derive from the automatically-adaptive temporal resolution in \textit{WaterLily.jl}, maintaining an approximately constant maximum CFL number. This causes small timing discrepancies in changes in acceleration, and thus forces, across spatial-resolution cases. Aside from this numerical artifact, the largest differences compared with the $N=128$ resolution case are roughly $8\%$ and $3\%$ for resolutions of $N=32$ and $N=64$, respectively. Based on this, we consider the case with $N=64$ sufficiently spatially-resolved for the purpose of this study, as the wake force contribution studied here is typically well above $3\%$. 

\subsection{Domain Size}
\label{subsec:appendix_domain}

\begin{figure}
    \subcaptionbox{\centering\hspace{-1.2cm}\label{fig:convergence_domain}}{\includegraphics[width=0.465\textwidth]{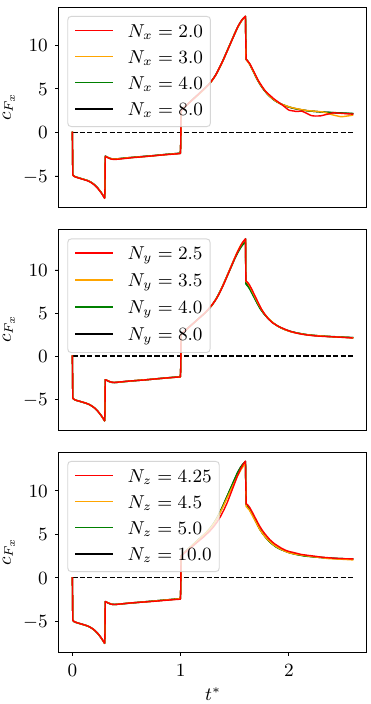}}%
    \hfill
    \subcaptionbox{\centering\hspace{-1.2cm}\label{fig:convergence_domain_percentage}}{\includegraphics[width=0.465\textwidth]{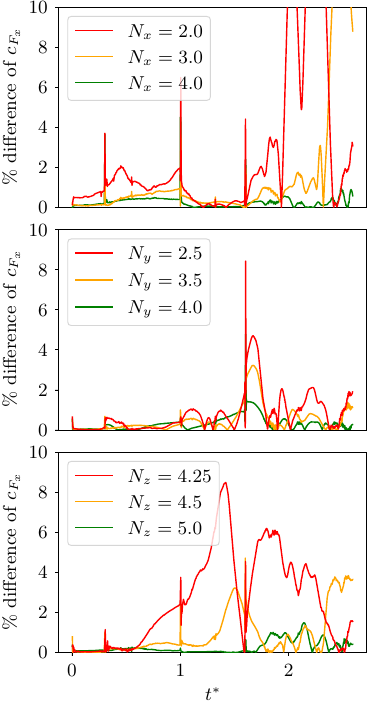}}%
    \hfill\null
    \caption{(a) Nondimensional force component in the $x$-direction, $\Cfx$, for varying spatial extent of the domain in $x$-, $y$- and $z$-direction, expressed in number of chord lengths as $N_x$, $N_y$ and $N_z$, respectively. The performed case has $\Rey=1000$ and $\dstar=1.0$. (b) Difference between the smaller domains and the largest domain case in each direction, expressed as a percentage of the instantaneous force from the largest domain case.}
    \label{fig:convergence_domain_all}
\end{figure}

To determine an appropriate numerical domain size, the lower acceptable limit in each orthogonal dimension was determined by iteratively reducing the spatial extent of the domain in that dimension while keeping the two other dimensions large. The plate is fixed at the centre of the computational domain and the prescribed motion is imposed through time-dependent inflow boundary conditions. The effect of domain reduction in each of the $x$-, $y$-, and $z$-directions on the forces computed on the plate can be seen in~\cref{fig:convergence_domain}. The differences between the force computed using each smaller domain and the largest domain are expressed as a percentage of the instantaneous force computed from the largest domain case, and shown in~\cref{fig:convergence_domain_percentage}. It is concluded that a domain size of $4.0c\times4.0c\times5.0c$ is sufficient for this case, as the differences in force resulting from domain size variation are less than those observed due to discretisation. For higher $\dstar$, the computational domain is extended in both the direction of plate motion ($x$) and along the plate chord ($y$) to accommodate the increased plate displacement and the growth of the vortex ring, respectively, with the vorticity fields inspected to verify that all vortical structures remain within the computational domain.

\subsection{Symmetry Boundary Condition}
\label{subsec:appendix_symmetry}

\begin{figure}
    \subcaptionbox{\centering\hspace{-1.2cm}\label{fig:convergence_symmetry}}{\includegraphics[width=0.465\textwidth]{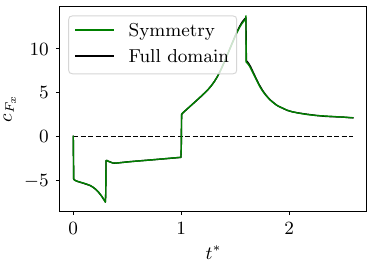}}%
    \hfill
    \subcaptionbox{\centering\hspace{-1.2cm}\label{fig:convergence_symmetry_percentage}}{\includegraphics[width=0.465\textwidth]{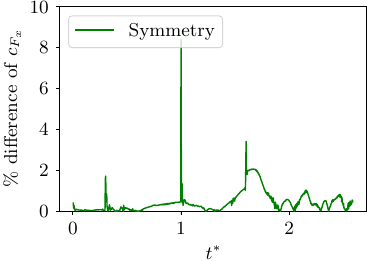}}%
    \hfill\null
    \caption{(a) Nondimensional force component, $\Cfx$, in the $x$-direction, with and without assuming $xy$- and $xz$- symmetry planes. The performed case has $\Rey=1000$ and $\dstar=1.0$. (b) Difference between the symmetry case and the full domain case, expressed as a percentage of the instantaneous force from the full domain case.}
    \label{fig:convergence_symmetry_all}
\end{figure}

The flat-plate geometry and prescribed kinematics are symmetric about both the $xy$- and $xz$-planes. The discretisation and domain-size studies in~\cref{subsec:appendix_discretisation,subsec:appendix_domain} also produced flow fields that were symmetric about these planes. To verify the validity of imposing symmetry boundary conditions, a simulation on a single $yz$-quadrant is compared with a full-domain simulation in~\cref{fig:convergence_symmetry}. Both simulations use the discretisation described in~\cref{subsec:appendix_discretisation}. The full-domain simulation uses the computational domain introduced in~\cref{subsec:appendix_domain}, while the quadrant simulation uses one quarter of this domain, bounded by the $xy$- and $xz$ symmetry planes. The instantaneous percentage difference between the forces obtained from the single-quadrant and full-domain simulations, computed relative to the corresponding force from the full-domain simulation, is shown in~\cref{fig:convergence_symmetry_percentage}. The observed differences are smaller than the errors introduced from discretisation. The symmetry assumption is therefore accepted and used for all simulations presented in the main text.

\section{Experimental Validation}
\label{sec:appendix_expval}

\renewcommand{\thefigure}{B\arabic{figure}}
\setcounter{figure}{0}

\begin{figure}
    \centering
    \hfill
    \subcaptionbox{\centering\hspace{-1.2cm}\label{fig:exp_motion}}{\includegraphics[width=0.5\textwidth]{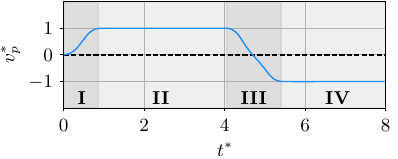}}%
    \hfill
    \subcaptionbox{\centering\label{fig:exp_setup}}{\includegraphics[width=0.5\textwidth]{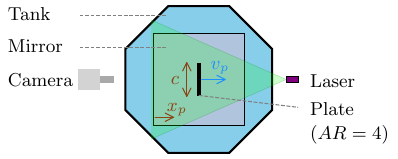}}%
    \hfill\null
    \caption{(a) Kinematics of the flat plate forward translation and reversal motion experimentally realised here: Dimensionless velocity, $\vpstar$, over time, for the case in which reversal occurs at $\dstar=4$. Roman numerals delineate distinct phases of the motion history: \textbf{I}. linear acceleration, \textbf{II}. translation at constant forward velocity, $\vpstar$, in $\xstar$, \textbf{III}. linear deceleration, \textbf{IV}. translation at constant reverse velocity, $-\vpstar$. (b) Schematic of experimental setup, viewed from above, showing location and motion direction of the $AR=4$ flat plate within the an octagonal, fluid-filled glass tank. Direction of illumination via laser sheet is indicated. The camera, used for digital PIV, views the region of interest through a mirror, mounted at an angle underneath the tank.}
    \label{fig:exp_kinematics_setup}
\end{figure}

To validate the numerical findings, a series of supporting experiments was conducted using the same flat-plate geometry as in the numerical simulations, with physical dimensions of $\PlateShape$ and a translation velocity of $\vp=\SI{0.2}{\meter\per\second}$. The plate was mounted on a six-axis robotic arm (\RobotType) using a $\SI{200}{\milli\meter}$ long cylindrical strut with a diameter of $\SI{15}{\milli\meter}$. The prescribed motion closely reproduced the numerical translation--reversal kinematics shown in~\cref{fig:schematic_kinematics}. Due to experimental constraints, minor deviations from the prescribed numerical motion were unavoidable, primarily during the acceleration phase; the measured experimental kinematics are therefore provided in~\cref{fig:exp_motion}, and the experimental validation remains largely qualitative. The experiments were performed in an octagonal glass tank (see~\cref{fig:exp_setup}) with side length $\TankSide$ and fluid depth $\WaterDepth$, filled with a water--glycerol mixture with glycerol volume fraction $\VolumeFractionGlycerol$, density $\rho=\Density$, and dynamic viscosity $\mu=\Viscosity$. These experimental conditions resulted in a Reynolds number of $\Rey=1165$, similar, but not identical, to the numerical case which produced the flow presented in~\cref{fig:Q_Re1000} and~\cref{fig:Re1000_fields_xy}.

The forces acting on the plate were measured using a force sensor (\ForceSensor) at a sampling frequency of $f_f=\ForceFrequency$. Simultaneously, the flow field in the plate $xy$-aligned midplane was measured using two-component, two-dimensional planar Particle Image Velocimetry (PIV) (\SettingsDavis). The flow was seeded with tracer particles (\PIVParticleType; 
density $\PIVParticleDensity$), and the measurement plane was illuminated by a \LaserType{} laser (\LaserSpecifications). Images were acquired using a \CameraBrand\ \CameraModel\ camera operating at a frame rate of $f_c=\CameraFrequency$, with a sensor resolution of $\CameraResolution$ pixels, a pixel size of $\CameraPixelSize$, and a field of view of $\FieldOfView$. The laser sheet thickness was approximately $\PIVLaserSheetThickness$. The imaging lens was operated at an aperture of $f/\CameraFStop$, giving a depth of field of well beyond the laser sheet thickness.

\begin{figure}
    \centering
    \includegraphics[width=\linewidth]{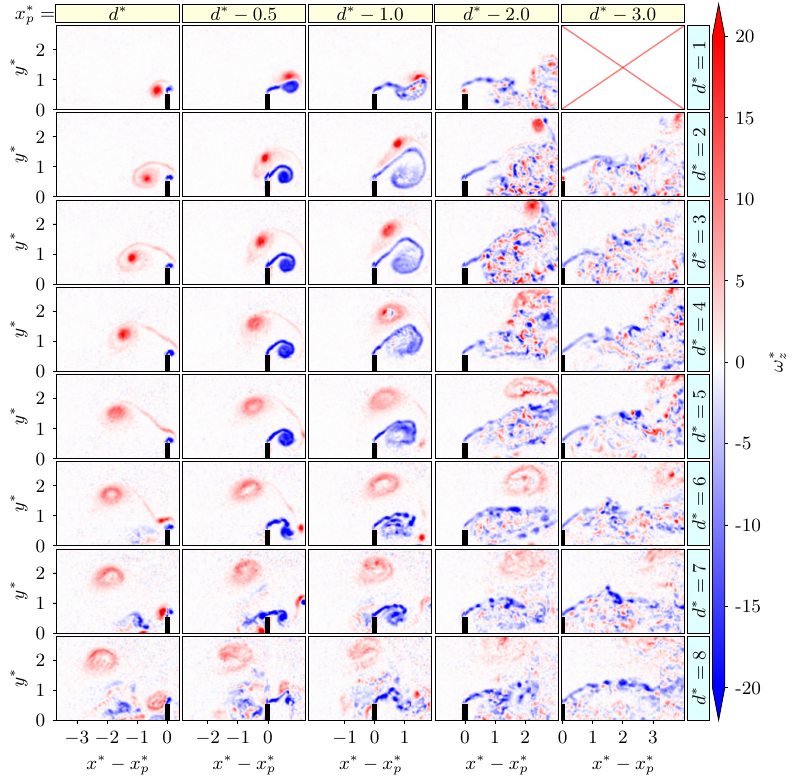}
    \caption{Spanwise-component, $\omegazstar$, vorticity fields from experiment at $\Rey=1165$ in the plate $xy$-midplane at selected phases $\dstar$, $\dstar-0.5$, $\dstar-1.0$, $\dstar-2.0$, and $\dstar-3.0$ throughout the post-reversal motion and for various pre-reversal distance cases, $\dstar\in[1,8]$. The top half of symmetric fields are shown, for brevity. A panel containing a red cross indicates data which were not recorded.}
    \label{fig:exp_vorticity_fields_Re1000}
\end{figure}

\begin{figure}
    \centering
    \includegraphics[width=0.5\linewidth]{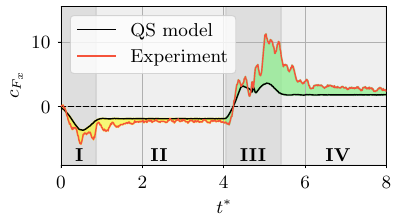}
    \caption{Force $x$-component over time, from experiment at $\Rey=1165$, $\dstar=4.0$. The red solid curve is computed from numerical simulation. The black dashed curve is predicted by the QS model, based on the measured plate motion. Roman numerals delineate the phases of the plate's motion: \textbf{I}. linear acceleration, \textbf{II}. constant forward velocity, \textbf{III}. linear deceleration, \textbf{IV}. constant reverse velocity. Discrepancy between computation and prediction is shaded: yellow during forward motion, and green during reverse.}
    \label{fig:exp_cd_QS_veff_ds4}
\end{figure}

\begin{figure}
    \centering
    \includegraphics[width=\linewidth]{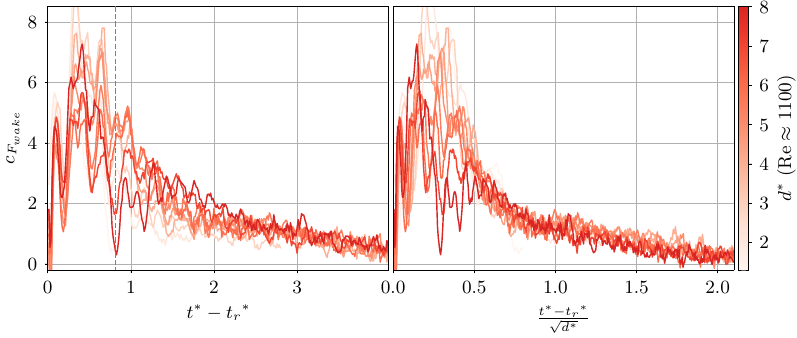}
    \caption{Wake interaction force, $\cdwake$, from experiments, over time since reversal (left) and over post-reversal time scaled by $\sqrt{\dstar}$ (right). Dashed vertical line indicates the moment at which deceleration stops.}
    \label{fig:exp_cd_wake_all}
\end{figure}

Experimental $z$-component vorticity fields in the $xy$-midplane for the same cases and at the same moments as for the numerical simulations~(see~\cref{fig:Re1000_fields_xy}) are plotted in~\cref{fig:exp_vorticity_fields_Re1000}. The main flow features are much like in the results of the numerical simulations, with vortex formation, pinch-off, and re-formation following the same progression. Both the lateral and streamwise trajectories of the initial starting vortex and reverse starting vortex match those observed in the numerical simulations. Smaller-scale structures are also qualitatively similar between simulation and experiment, for example the formation of a flow structure to the left of the plate at $\dstar\gtrsim 6$ in the left column. At late times, post-reversal, small-scale structures are no longer highly similar between simulation and experiment, most likely pointing to a chaotic, turbulent, origin of these structures.

Phase-averaged measurements of the force coefficient in the $x$-direction are shown in~\cref{fig:exp_cd_QS_veff_ds4}~(red curve) for a motion case with $\dstar=4.0$. These can be compared to the numerical case in~\cref{fig:cd_QS_veff_ds4}. A QS estimate of the force based on kinematics retrieved from the robot arm is included for comparison (black curve). Here, as in the numerical case, a wake interaction force is present during the reversal-phase of the plate's motion, indicated by the green shading. This wake interaction force peaks late in the deceleration, with a magnitude similar to that numerically computed, then monotonically decays in time after the deceleration completely stopped.

The wake interaction force was isolated by subtracting the QS estimate from the experimentally measured force values, in the same way as was done for the numerical simulations. This was repeated for various $\dstar\in[1,8]$. The resulting force coefficients are given in~\cref{fig:exp_cd_wake_all}, along with an empirical scaling with $\sqrt{\dstar}$, showing similar peak, decay, and collapse. Despite some experimental noise, larger $\dstar$ cases demonstrate a similar double peak in the wake interaction force to what was observed in the numerical simulations.

\bibliographystyle{plainnat}
\bibliography{Mendeley}

\end{document}